\documentclass[fleqn,usenatbib]{mnras}

\usepackage{mathptmx}

\usepackage[T1]{fontenc}
\usepackage{ae,aecompl}

\usepackage{graphicx}	
\usepackage{amsmath}	
\usepackage{amssymb}	
\usepackage{color, soul}

\usepackage{hyperref}
\usepackage{cleveref}

\title[The highest-frequency kHz QPOs in NS-LMXBs]{The highest-frequency kHz QPOs in neutron star low mass X-ray binaries }

\author[van Doesburgh et al.]{
Marieke van Doesburgh,$^{1}$\thanks{E-mail: m.j.vandoesburgh@uva.nl}
Michiel van der Klis,$^{1}$
Sharon M. Morsink$^{2}$
\\
$^{1}$Anton Pannekoek Institute, University of Amsterdam, Science Park 904, Postbus 94249, 1090 GE Amsterdam, The Netherlands\\
$^{2}$Department of Physics, University of Alberta, CCIS 4-181, Edmonton, AB T6G 2E1, Canada\\
}

\date{Accepted XXX. Received YYY; in original form ZZZ}

\pubyear{2018}

\begin{document}
\label{firstpage}
\pagerange{\pageref{firstpage}--\pageref{lastpage}}
\maketitle

\begin{abstract}
We investigate the highest-frequency kHz QPOs previously detected with RXTE in six neutron star (NS) low mass X-ray binaries.
We find that the highest-frequency kHz QPO detected in 4U\ 0614+09 has a 1267 Hz 3$\sigma$ confidence lower limit on its centroid frequency. This is the highest such limit reported to date, and of direct physical interest as it can be used to constrain QPO models and the supranuclear density equation of state (EoS). 
We compare our measured frequencies to maximum orbital frequencies predicted in full GR using models of rotating neutron stars with a number of different modern EoS and show that these can accommodate the observed QPO frequencies. Orbital motion  constrained by NS and ISCO radii is therefore a viable explanation of these QPOs. In the most constraining case of 4U 0614+09 we find the NS mass must be M$<$2.1 M$_{\odot}$. From our measured QPO frequencies we can constrain the NS radii for five of the six sources we studied to narrow ranges ($\pm$0.1--0.7 km) different for each source and each EoS.

\end{abstract}




\section{Introduction}
Quasi-periodic oscillations (QPOs) in the X-ray emission from accreting neutron star low mass X-ray binaries (NS-LMXBs) have been observed with best-fit centroid frequencies up to $\nu_0\sim$1250 Hz (references are given below). 

As orbital frequencies of $\sim$kHz are expected in the close vicinity of neutron stars, QPOs at such frequencies have often been associated with orbital motion in the inner accretion disk (see for example \citealt{Stella:1998}, \citealt{Torok:2016}, \citealt{Kluzniak:2001}, \citealt{Miller:1998} and \citealt{Titarchuk:1999}).  
Additional evidence for this interpretation was presented by \cite{Bult:2015} in the form of a dependence of the pulse amplitude in the accreting millisecond pulsar SAX\ J1808.4--3658 on the kHz QPO frequency, indicating an azimuthal asymmetry is at the base of the QPO mechanism. Alternative physical explanations of kHz QPOs invoke for instance disc oscillation modes, see \cite{Avellar:2018} and \cite{Rezzolla:2003}. \\

The accretion disk is truncated at or above the neutron star surface, and for this reason, if the QPO frequency is the general relativistic orbital frequency at the inner disk edge ($\nu_{orb}$), the maximum QPO frequency puts limits on the mass (ISCO radius smaller than orbital radius) and radius (smaller than orbital radius) of the neutron star to first order in $j$ \citep{Miller:1998}:

\begin{align}
\emph{M}\leq\text{2.2}(\nu_{orb}\text{/}\text{1000 Hz})^{-1}(\text{1+0.75}\emph{j})\emph{M}_{\odot}
\label{eq:M}
\end{align} 
and 
\begin{align}
\emph{R}\leq\text{19.5}(\nu_{orb}\text{/}\text{1000 Hz})^{-1}(\text{1+0.2}\emph{j}) \text{km}.
\label{eq:R}
\end{align}

Here, $j\equiv cJ/GM^2$ is the dimensionless angular momentum (spin parameter) of the neutron star, with $J$ the angular momentum and $M$ the mass \citep{Miller:1998a}. Clearly, these constraints tighten when $\nu_{orb}$ increases. For this reason the question what is the highest reliably detected QPO frequency is of direct physical interest.

The highest kHz QPO frequency reported in the literature is $\nu_u$=1329$\pm$4 Hz (3.5$\sigma$, Q=51
, single kHz QPO) in 4U\ 0614+09 \citep{vanStraaten:2000}. In that work, a 1273$\pm$15 Hz (no significance quoted, Q=14
, single kHz QPO) QPO is also reported from another observation of the same source. Subsequent analysis by \cite{Boutelier:2009} (using an automated detection method with a lower limit on the detection significance of 6$\sigma$) of an extended data set on 4U\ 0614+09 including the observations in which the high-frequency kHz QPOs were reported by \cite{vanStraaten:2000} failed to confirm the presence of a significant kHz QPO $>$1224 Hz (the highest frequency is $\nu_0$=1224$\pm$6, Q=21 at $\sim$8.6$\sigma$). For the observation in which \cite{vanStraaten:2000} measured the 1329$\pm$4 Hz QPO the authors reported a $\sim$2.6$\sigma$ 1328$\pm$27 Hz QPO (Q=29, single kHz QPO). 
We list these detections in Table \ref{tab:records} along with the other highest-frequency kHz QPOs previously reported for other sources.

All sources in the table are atoll sources at low luminosity.  In all cases the high-frequency kHz QPO was detected when the source was in the banana state \citep{vanStraaten:2000, Jonker2007, Kaaret:2002, Altamirano:2008, DiSalvo:2001, Markwardt:1999}. Burst oscillations were detected in other observations of 4U\ 0614+09, SAX\ J1750.8--2900, 4U\ 1636--53, 4U\ 1702--43 and 4U\ 1728--34; the NS spin frequency is known for these sources. SAX\ J1750-2900 and 4U\ 1636--53 were included in the study by \cite{vandoesburgh:2016}, but no higher frequencies than reported in the literature were found.  In \cite{Jonker:2001} it was suggested that the fractional rms of the kHz QPO is anti-correlated with source luminosity. All sources mentioned here seem to follow this trend, indicating that the high frequency QPO states of these varied sources might form a homogeneous group.

\begin{table*}
\centering
 \tabcolsep=0.1cm
 {
\scalebox{1}{
\small 

	\begin{tabular}{l c c c c c c c} 
		\hline
		Source & Spin & ObsID &$\nu_0$ & Q & Sign.  & single/twin &  ref.\\
		       & (Hz)&  & (Hz)  & & (IP/$\sigma^-$) & &  \\
		\hline
		4U\ 0614+09 & 415 & 40030-01-04-00 & 1329$\pm$4 & 51 &  3.5 & single & \cite{vanStraaten:2000} \\
		4U\ 0614+09 & & 30056-01-03-03(+2 obs) & 1273$\pm$15 & 14 &  3.5 & single & \cite{vanStraaten:2000} \\
		4U\ 0614+09 & & 30056-01-05-00 & 1224$\pm$6 & 21 & 8.6 & single & \cite{Boutelier:2009} \\
		4U\ 0614+09 & & 40030-01-04-00 & 1328$\pm$4 & 29 &  2.6 & single & \cite{Boutelier:2009} \\
		4U\ 0614+09 & & 40030-01-06-00 & 1369$\pm$52 & 4 &  3.9 & single & \cite{vandoesburgh:2016} \\
		            \hline
		            
		4U\ 1246--59 & unknown &90042-02-08-00 & 1258$\pm$4 & 50 & $>$7 & single & \cite{Jonker2007} \\ 
		\hline
		SAX\ J1750--2900 & 601 & 60035-01-01-00 & 1253$\pm$9 & 20 & $>$5 & twin & \cite{Kaaret:2002} \\ 
		\hline
		4U\ 1636--53 &581 & 60032-01-03-01(+30 obs) & $\nu_{max}$=1259$\pm$10 & 14  & n.a. & single & \cite{Altamirano:2008} \\
		\hline
		4U\ 1728--34 & 363 & selection, see ref. & 1161$\pm$16 & 9 & $>$4 & twin & \cite{DiSalvo:2001} \\
		4U\ 1728--34 & & 20083-01-02-000/01 & 1276$\pm$59 & 5 & 2.89 & single & \cite{vandoesburgh:2016} \\
		\hline
		4U\ 1702--43 & 329 &20084-02-01-02/020 & 1155$\pm$5 & 11 & $>$5 & twin & \cite{Markwardt:1999}\\
		4U\ 1702--43 & & 80033-01-10-00(+10) & 1198$\pm$15 & 13 & 3.7 & single & \cite{vandoesburgh:2016} \\
	\end{tabular}	
	}}
	\caption{The highest kHz QPOs reported in the literature. As indicated, \protect\cite{Altamirano:2008} used fit parameter $\nu_{max}$ instead of centroid frequency ($\nu_0$) for 4U\ 1636-53 ($\nu_{max}=\nu_0\sqrt{1+\frac{1}{4Q^2}}$). }	
	\label{tab:records}
\end{table*}

In \cite{vandoesburgh:2016} we recently reported a highest kHz QPO frequency in RXTE archival data on 4U 1728--34 of 1276$\pm$38 Hz, on 4U\ 0614+09 of 1369$\pm$51 Hz and on 4U\ 1702--43 of 1198$\pm$15 Hz. In that work, a large data set was uniformly processed for a different purpose than identifying the highest QPO frequencies. In view of the relevance of the issue of the highest frequency reliably detected for a kHz QPO, in this paper we present a detailed study of the QPO frequencies reported in \cite{vandoesburgh:2016}, compare these to the literature (reanalyzing the data when necessary) and discuss implications for the supranuclear density equation of state and kHz QPO models.\\ \\

\section{Data analysis}
\subsection{Energy spectral analysis}
We use the Crab-normalized soft and hard colours averaged per RXTE observation (single pointing of $\sim$1.5--2.5 ks in length, archived under a single ObsID) as previously obtained by \cite{vandoesburgh:2016} following the method of \cite{vanStraaten:2000}. The hard colour is defined as the ratio between the counts in the 9.7--16.0 keV and 6.0--9.7 keV energy bands and the soft colour as that between the 3.5--6.0 keV and 2.0--3.5 keV bands.

\subsection{Timing analysis}
To calculate the power spectra we use GoodXenon and Event mode data with a time resolution of 1/8192 s ($\sim$122 $\mu$s) or better. We rebin if necessary to 1/8192 s, and divide the data into segments of 16 seconds. This results in power spectra with a Nyquist frequency of 4096 Hz and a lowest frequency and frequency resolution of 0.0625 Hz. 
We include channels corresponding to photon energies of 2 to 18 keV for reasons outlined in Section \ref{sec:energy}. 
No background or dead time corrections were performed before calculating the power spectra; instead, we correct for these effects after averaging the Leahy-normalized power spectra. To do so, we subtract a counting noise model power spectrum incorporating dead-time effects \citep{Zhang:1995} following the method of \cite{Klein:2004PhD}.  We renormalize the power spectra such that the square root of the integrated power in the spectrum equals the fractional root mean square amplitude (rms) of the variability in the signal \citep{Klis:1989}.

\subsubsection{Energy channel selection}
\label{sec:energy}
\begin{figure}
	\includegraphics[width=3.5in]{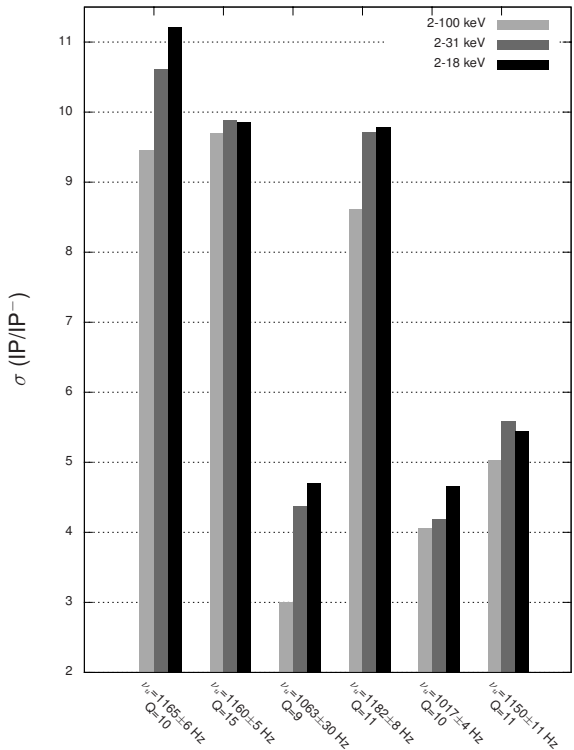}
   \caption{Detection significance in different energy ranges of upper kHz QPOs (with $\nu_u$>1000 Hz). ObsIDs used, from left to right: 30056-01-01-00, 80037-01-06-03, 80037-02-01-00, 40429-01-03-06, 30056-01-04-01 (all 4U\ 0614+09), 95337-01-02-00/000 (4U\ 1728--34).}
   \label{fig:obs_kHz}
\end{figure}

To determine the optimal energy band for the detection of the upper kHz QPO with $\nu_u$>1200 Hz, we obtain the detection significance of a selection of upper kHz QPOs reported in \cite{vandoesburgh:2016} at slightly lower frequency (1000$<\nu_u<$1200 Hz) in 3 different energy ranges: 2-18 keV, 2-31 keV (we correct for detector gain changes over time by selecting the appropriate channels), and all available energy channels (here, the maximum energy can vary between epochs).
To increase signal to noise, we average the 16 second power spectra within an observation.
We fitted the power spectra above 400 Hz with a constant (to account for the Poisson noise level, fixed to zero as we subtracted it in previous steps),  plus two Lorentzians (all Lorentzian parameters are allowed to vary): the lower and upper twin kHz QPOs (designated L$_{\ell}$ and L$_u$, respectively).
We find that this frequency selection does not affect the frequency or the detection significance of the kHz QPOs compared to fitting the entire frequency range with a more complex model. Also, alternative handling of the Poisson noise level by using a floating constant did not affect the frequency or detection significance of the kHz QPOs. 
In Figure \ref{fig:obs_kHz} we plot the detection significance of L$_u$, approximated by the ratio of the best-fit integrated power in the power spectral component to its negative 1$\sigma$ error, for different energy ranges. We find that, overall, a 2-18 keV energy selection optimizes the detection significance of L$_u$ in the 1000$<\nu_u<$1200 Hz frequency range.

 We apply the 2-18 keV energy selection to construct power spectra for the observations in which the $\nu_u>1200$ Hz QPOs were reported. A further cut in energy suggested by the slightly lower hard color of these observations does not improve the detection significance further.

\subsection{Data selection}

\begin{figure*}
	\includegraphics[width=4.5in]{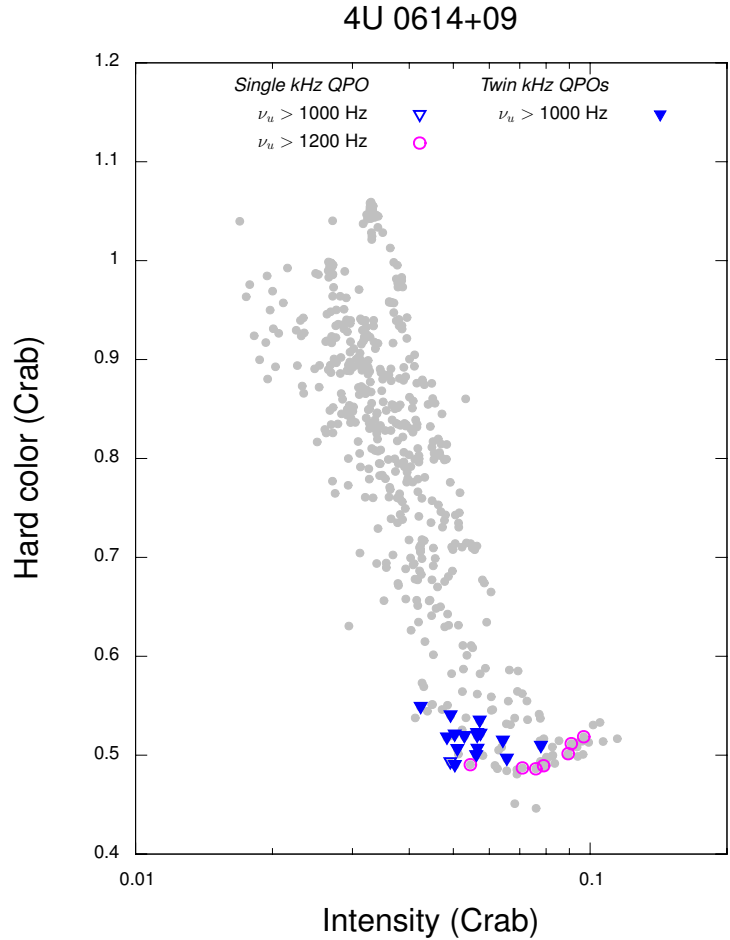}
   \caption{Crab-normalized hard color vs. intensity averaged for each observation in the RXTE archive of 4U\ 0614+09. The ObsIDs in which we detected kHz QPOs with $\nu_u>$1200 Hz in \protect\cite{vandoesburgh:2016} are indicated by pink open circles. Error bars are the size of the symbols.}
   \label{fig:HI}
\end{figure*}

\begin{figure*}
	\includegraphics[width=4.5in]{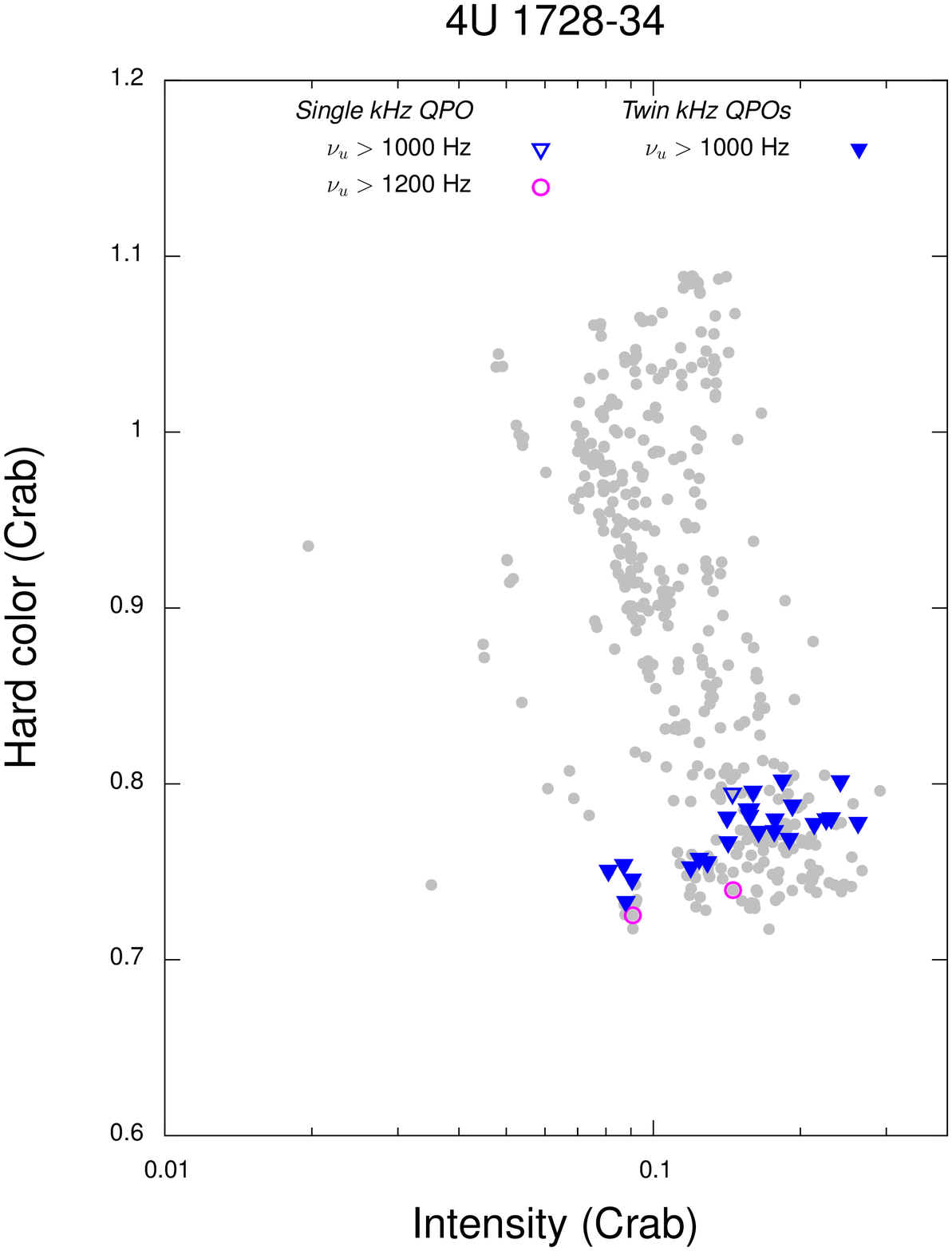}
   \caption{Same as Figure \ref{fig:HI}, for 4U\ 1728--34.}
   \label{fig:HI2}
\end{figure*}

We plot hard colour vs. intensity, per observation, for 4U\ 1728-34 and 4U\ 0614+09 in Figures \ref{fig:HI} and \ref{fig:HI2}. We indicate the observations for which twin kHz QPOs and single kHz QPOs were reported with $\nu_u$>1000 Hz in \cite{vandoesburgh:2016}. The observations with $\nu_u$>1200 Hz (pink) all have single kHz QPOs and low hard color. The frequency of the upper kHz QPO when it appears in a pair of twin QPOs evolves smoothly to the frequencies of the QPO detected as a single peak toward lower hard color. We therefore identify it as the upper kHz QPO.

In an attempt to detect other high-frequency kHz QPOs, we fitted observations, both single and averaged, with high frequency QPOs. For 4U\ 0614+09 we focused on observations with hard colour $<$0.5 or intensity $>$0.08 Crab, and for 4U\ 1728--34 on observations with hard color $<$ 0.76. We detect a >3$\sigma$ 1269$\pm$12 Hz QPO in 4U\ 0614+09 (in 96404-01-10-01) not reported in previous works. No other kHz QPOs were found. 


\subsection{Power spectral fitting}

 We fit the power spectra $>400$ Hz as described in section \ref{sec:energy}. As the FWHM of the upper kHz QPOs in 4U\ 0614+09 and 4U\ 1728-34 above 1000 Hz is typically 50-150 Hz \citep{Boutelier:2009, vandoesburgh:2016}  we apply logarithmic binning with 100 bins per decade so that the QPO peak is covered by a sufficient number of frequency bins without over-resolving it. Although subtly, the binning factor can affect the detection significance, as can be seen in Table \ref{fit}, where for comparison purposes we also list fits to power spectra binned with 75 bins per decade.

\section{Results}
In Table \ref{fit} we list the best-fit parameters of fits to power spectra obtained when including all energy channels and fitting all frequencies (the method we used in \citealt{vandoesburgh:2016}) as well as those to power spectra obtained in the 2-18 keV energy range and only fitting frequencies $>$400 Hz. These results confirm the expectation based on the $<$1200 Hz QPO fits that the signal to noise generally improves when selecting the 2-18 keV energy band (this is the case in 13 out of 17 cases). None of the three fit parameters (centroid frequency, FWHM and integral power) significantly change for the repeated fits reported in the Table. We detect single upper kHz QPOs (L$_u$) in 16 out of 17 cases. Below, we discuss Table \ref{fit} in further detail. 

\subsection{4U\ 0614+09}

The highest frequency QPO detected at better than 3$\sigma$ significance, a $>$5$\sigma$ QPO in 30056-01-03-03, has $\nu_u$=1288$\pm{8}$ Hz. 
This is within 1$\sigma$ of the 1273$\pm{15}$ Hz QPO reported in \cite{vanStraaten:2000}, which was measured in an average of 30056-01-03-03 with two other observations close in time.\\
A number of other observations merit mentioning.
 In 40030-01-04-00, the observation previously suspected to feature the highest frequency QPO, we detect a 1311$^{+34}_{-57}$ Hz QPO (2.5$\sigma$) with Q$\sim$9. This is a broader QPO than the Q$\sim$29, 1328$\pm$27 Hz one ($\sim$2.6$\sigma$) reported by \cite{Boutelier:2009}. We do not reproduce the Q$\sim$50, 1329$\pm{4}$ Hz (3.5$\sigma$) QPO reported by \cite{vanStraaten:2000}. 

In \cite{vandoesburgh:2016} we reported a 1369$\pm$51 Hz QPO for 40030-01-06-00. In our current analysis we detect a 4.5$\sigma$ QPO with a centroid frequency of 1283$^{+43}_{-39}$ Hz. The frequency difference between the two results is not significant (1.4$\sigma$). 
 To further improve signal to noise, we average 40030-01-04-00 and 40030-01-06-00, as these observations are close in time, have similar hard and soft color (to within 10$\%$) and show similar power spectra. We obtain a 1285$\pm{31}$ Hz QPO at 4.9 $\sigma$. 
 
In 40030-01-05-00, twin kHz QPOs are detected, with the upper kHz QPO at 1333$\pm$25 Hz (2.3$\sigma$). The two QPOs (the lower peak is at 1121$\pm{16}$ Hz, 3$\sigma$) have a peak separation of 212$\pm{40}$ Hz, which is slightly less than (but still within 2$\sigma$ of) the 320$\pm40$ Hz seen for $\nu_u<$1170 Hz \citep{Boutelier:2009}, so the two peaks may move together when the frequencies increase, as expected.

\subsection{4U\ 1728--34}
For 4U\ 1728--34, we exceed the previously reported highest frequencies of 1161$\pm$16 Hz (detected in an ensemble of observations with similar energy spectra) and 1276$\pm$59 (detected in a combination of 20083-01-02-000 and 20083-01-02-01, see Table \ref{tab:records}).
We detect a highest QPO frequency of 1302$^{+13}_{-10}$ Hz (3.3 $\sigma$) in observation 20083-01-02-000 alone. 
 The best constrained frequency however, corresponds not to this 1302$^{+13}_{-10}$ Hz peak, but to the 1278$^{+15}_{-14}$ Hz one detected in the combination of observations 50030-03-09-00 and 50030-03-09-01; see Section \ref{sec:limits}. 
\subsection{4U\ 1636--53}
Because of its high spin frequency (581 Hz), which makes this source particularly important for constraining the equation of state (see section \ref{sec:eos}), we reanalyzed the high frequency kHz QPO reported for 4U\ 1636-53 by \cite{Altamirano:2008} for the ensemble of 31 observations in "Group N" listed in Table 6.1 of that work. We detect a 1255$^{+13}_{-22}$ Hz QPO at 4.4$\sigma$.
\subsection{SAX\ J1750.8--2900}
The highest kHz QPOs in SAX\ J1750.8--2900, another source with high spin (601 Hz), were reported by \cite{Kaaret:2002}.
Following the analysis in that work, we use all data of observation 60035-01-01-00. We detect a 1250$^{+9}_{-7}$ Hz QPO at 4.4 $\sigma$ which is consistent with the earlier report. 
\subsection{4U\ 1246--59}
For 4U\ 1246--59 (spin frequency unknown), \cite{Jonker2007} reported a 1258$\pm$2 Hz QPO using an average of three observations. For the same data, we find a 1261$^{+0.9}_{-1.3}$ Hz QPO when using the 2-18 keV energy selection. This is consistent with the result reported by \cite{Jonker2007}. Because the QPO is so narrow (Q$>$50), a high bin resolution is required to resolve it. This can be seen in Table \ref{fit}, where the parameters in the fit with 75 bins per decade are badly constrained. We therefore deviate from our standard of 100 bins per decade for this QPO and use 200 instead.
\subsection{4U\ 1702-43}
Recently, \cite{Nattila:2017} obtained radius measurements by modeling the X-ray burst cooling tail spectra of 4U\ 1702--43. The authors fit atmosphere models directly to the data and find that the radius is constrained to be R=12.4$^{+0.6}_{-2.6}$ at 97.7$\%$ confidence. As it is interesting to compare the radius constraint obtained with spectral modeling to the constraint of the maximum QPO frequency on radius, we include 4U 1702--43 in our source selection.  \cite{vandoesburgh:2016} report a highest kHz QPO of 1198$^{+16}_{-14}$ Hz in a combination of 11 observations when including all energy channels. In our current analysis we detect a 1213$^{+26}_{-17}$ Hz QPO.

\subsection{Lower limits}
\label{sec:limits}
\begin{figure*}
	\includegraphics[width=4.5in]{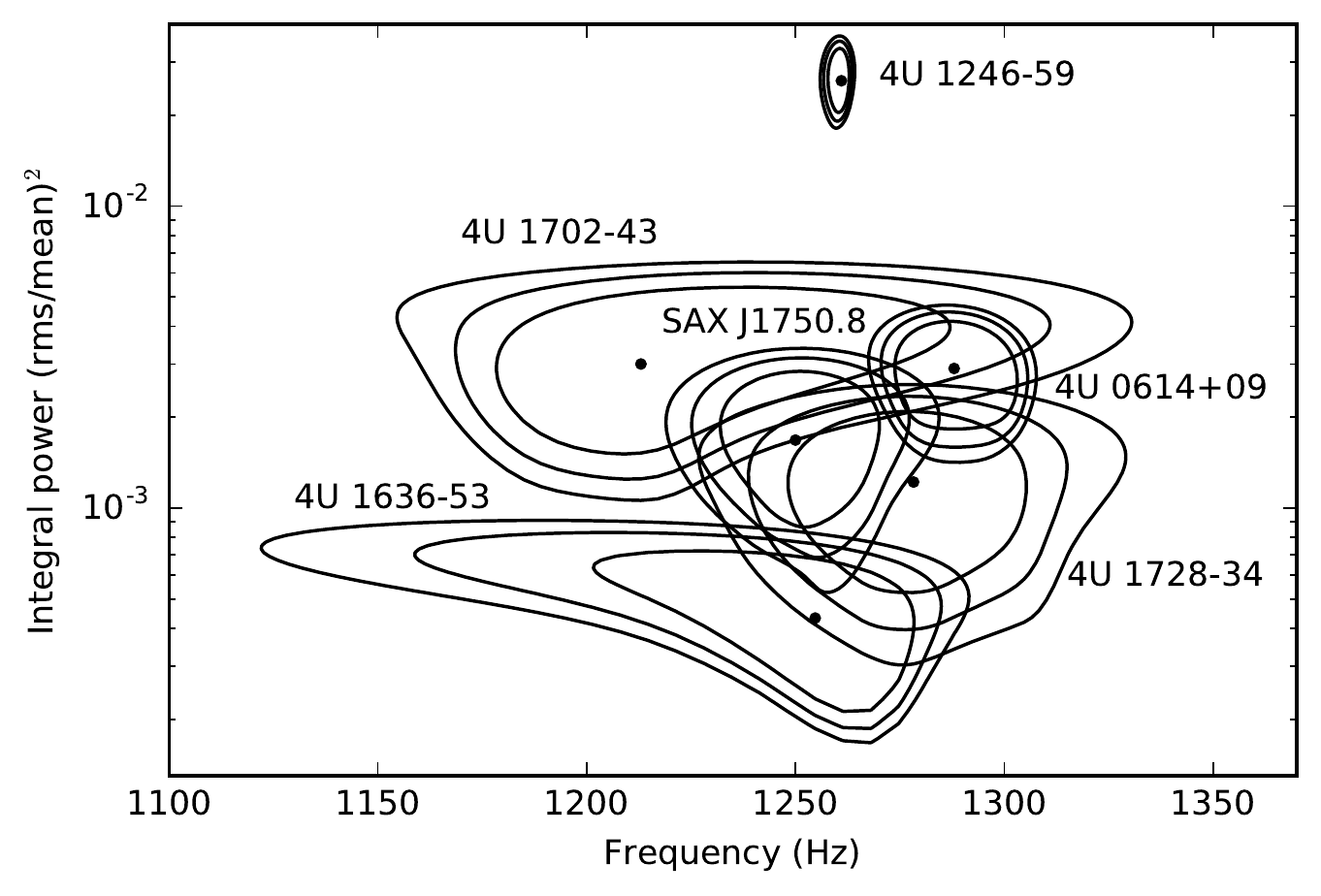}
   \caption{Confidence contours for the best-fit $\nu_u$ and integral power of the most confidently detected highest kHz QPOs in 4U\ 0614+09, 4U\ 1728--34, 4U\ 1636--53, SAX J1750--2900, 4U\ 1702-43 and 4U\ 1246--59. We plot the contours corresponding to $\Delta\chi^2$=4.00, 6.00 and 7.74. The 3$\sigma$ (99.73$\%$ confidence) lower limit corresponds to the lowest frequency covered by the outer contour.}
   \label{fig:cont}
\end{figure*}

For the purpose of using the highest-frequency QPOs to constrain the EoS, it is useful to obtain the highest observed lower limits on the QPO frequencies. 
The probability distributions of the QPO frequencies deduced from our multi-parameter fits are not generally strictly Gaussian. In the last column of Table \ref{fit} we list the 3$\sigma$ confidence lower limits on the frequencies obtained by evaluating the frequency at which $\chi^2$ exceeds the best-fit value by $\Delta\chi^2$=7.74, leaving all other fit parameters free. We find that in several cases the $\chi^2$ hypersurface is less curved, and hence the constraint on frequency is weaker than we would have expected from the 1$\sigma$-error in the Gaussian approximation. In Figure \ref{fig:cont} we plot the confidence contours for the best-constrained upper kHz QPO frequencies detected in 4U\ 1728--34, 4U\ 0614+09 , 4U\ 1636-53, SAX\ J1750--2900 , 4U\ 1246--59 and 4U\ 1702--43 listed in Table \ref{fit}. 
The kHz QPO that is best constrained to have a high frequency is the 1288$\pm$8 Hz one in observation 30056-01-03-03 of 4U\ 0614+09, which is detected at $>$5$\sigma$. The 3$\sigma$ lower limit on its centroid frequency is 1267 Hz.

\section{Discussion}
Our highest 3$\sigma$ lower limit on a kHz QPO frequency, in 4U\ 0614+09, is 1267 Hz. For 4U\ 1728--34, 4U\ 1636--53, SAX\ J1750.8--2900, 4U\ 1246--59 and 4U\ 1702--43 these values are 1221, 1119, 1219, 1256 and 1150 Hz, respectively.

 \begin{table*}
\centering
  \tabcolsep=0.1cm
 {
\scalebox{0.9}{
\small
  \begin{tabular}{l c c c c c c c c}

   \hline 
\vspace*{0.05cm}
 Source  & ObsID & $\nu_0$ & FWHM  & IP ($\times$10$^{-3}$) & $\sigma$ (IP/$\sigma_-$) & Log. bin. factor & Energy channel & 3$\sigma$ lower limit \\
 & & (Hz) & (Hz) & & & &  selection &(Hz)  \\
\hline
4U\ 0614+09 &  40429-01-06-00   &1220$^{+8}_{-6}$&35$^{+19}_{-15}$&4.8$^{+1.6}_{-1.5}$&3.2&  -100   &  All &  \\
     &   &1220$^{+13}_{-11}$&71$^{+41}_{-30}$&5.7$^{+1.9}_{-1.7}$&3.4&  -75 ($>$400 Hz) &  0-43 &  \\
             &   & \textbf{1200}$^{\textbf{+18}}_{\textbf{-20}}$&\textbf{104}$^{\textbf{+80}}_{\textbf{-40}}$&\textbf{6.2}$^{\textbf{+2.4}}_{\textbf{-1.9}}$&\textbf{3.3}& \textbf{-100}($>$\textbf{400} \textbf{Hz}) &  \textbf{0-43} &  \textbf{818}\\
\hline
 & 30056-01-05-00   &1224$\pm{5}$&50$^{+9}_{-8}$&6.7$\pm{1}$&6.9&  -100 &  All & \\
    &   &1233$^{+6}_{-7}$&49$^{+12}_{-17}$&6.6$^{+1.0}_{-0.9}$&7.3&  -75 ($>$400 Hz) &  0-49 &  \\
        &   &\textbf{1221}$^{\textbf{+5}}_{\textbf{-3}}$&\textbf{45}$^{\textbf{+12}}_{\textbf{-14}}$&\textbf{6.6}$^{\textbf{+0.9}}_{\textbf{-0.8}}$&\textbf{8.0}&  \textbf{-100} ($>$\textbf{400} \textbf{Hz}) &  \textbf{0-49} &  \textbf{1214} \\ 
\hline
 & 30056-01-03-04  &1285$^{+27}_{-25}$&120$^{+89}_{-48}$&2.8$^{+1.1}_{-0.9}$&2.9&  -100 &  All &  \\ 
     &            &1285$^{+43}_{-30}$&159$^{+112}_{-88}$&2.9$^{+1.2}_{-1.0}$&2.9&  -75($>$400 Hz)  &  0-49 & \\
     &            &\textbf{1292}$^{\textbf{+28}}_{\textbf{-27}}$&\textbf{133}$^{\textbf{+100}}_{\textbf{-55}}$&\textbf{2.8}$^{\textbf{+1.1}}_{\textbf{-0.9}}$&\textbf{2.9}&  \textbf{-100}($>$\textbf{400} \textbf{Hz})  &  \textbf{0-49}&  \textbf{0} \\
\hline
 &  30056-01-03-03  &1286$^{+8}_{-10}$&50$^{+15}_{-17}$&2.6$\pm{0.6}$&4.6&  -100 &  All &  \\
  &   &1284$^{+5}_{-4}$&44$^{+28}_{-117}$&3.0$^{+0.6}_{-0.4}$&6.8&  -75 ($>$400 Hz) &  0-49 &  \\
  &   &\textbf{1288}$^{\textbf{+9}}_{\textbf{-8}}$&\textbf{61}$^{\textbf{+16}}_{\textbf{-13}}$&\textbf{2.9}\textbf{$\pm$0.6}&\textbf{5.3}&  \textbf{-100} ($>$\textbf{400} \textbf{Hz}) &  \textbf{0-49} &  \textbf{1267} \\
  \hline
     & 40030-01-05-00  L$_u$  &1328$^{+15}_{-17}$&71$^{+58}_{-46}$&2.6$^{+1.2}_{-1.0}$&2.7&  -100 &  All &   \\
   &  L$_u$ &1321$^{+33}_{-27}$&128$^{+151}_{-86}$&2.8$^{+1.9}_{-1.3}$&2.1&  -75 ($>$400 Hz) &  0-49 &  \\  
     &  L$_u$ &\textbf{1333}$^{\textbf{+24}}_{\textbf{-25}}$&\textbf{93}$^{\textbf{+92}}_{\textbf{-56}}$&\textbf{2.4}$^{\textbf{+1.4}}_{\textbf{-1.0}}$&\textbf{2.3}&  \textbf{-100} ($>$\textbf{400} \textbf{Hz}) &  \textbf{0-49} &  \textbf{0}\\     
\hline
   &  L$_{\ell}$ &1112$^{+21}_{-16}$&71$^{+51}_{-40}$&2.4$^{+1.1}_{-0.9}$&2.6&  -100 &  All &  \\
   &  L$_{\ell}$ &1111$^{+21}_{-15}$&72$^{+49}_{-58}$&2.4$^{+1.1}_{-1.0}$&2.3&  -75 ($>$400 Hz) &  0-49 &   \\  
      &  L$_{\ell}$  &\textbf{1121}$^{\textbf{+16}}_{\textbf{-15}}$&\textbf{72}$^{\textbf{+37}}_{\textbf{-26}}$&\textbf{2.7}$^{\textbf{+1.0}}_{\textbf{-0.9}}$&\textbf{3.0}&  \textbf{-100} ($>$\textbf{400} \textbf{Hz}) &  \textbf{0-49} &  \textbf{n.a.} \\
\hline
 & 40030-01-04-00  &1318$^{+24}_{-65}$&108$^{+134}_{-76}$&1.8$^{+1.1}_{-0.9}$&2.1&  -100 &  All &   \\ 
 &   &1284$\pm{29}$&125$^{+84}_{-55}$&2.1$^{+0.8}_{-0.7}$&2.8&  -75 ($>$400 Hz) &  0-49 &  \\
  &  &\textbf{1311}$^{\textbf{+34}}_{\textbf{-57}}$&\textbf{150}$^{\textbf{+135}}_{\textbf{-97}}$&\textbf{2.0}$^{\textbf{+1.0}}_{\textbf{-0.8}}$&\textbf{2.5}&  \textbf{-100} ($>$\textbf{400} \textbf{Hz}) &  \textbf{0-49} &  \textbf{917} \\
\hline

 & 40030-01-06-00  &1369$^{+52}_{-51}$&321$^{+154}_{-101}$&6.5$^{+1.9}_{-1.7}$&3.9&  -100 &  All &  \\
 &   &1287$^{+44}_{-42}$&331$^{+138}_{-92}$&6.2$^{+1.5}_{-1.4}$&4.5&  -75 ($>$400 Hz) &  0-49 &  \\
 &   &\textbf{1283}$^{\textbf{+43}}_{\textbf{-39}}$&\textbf{318}$^{\textbf{+135}}_{\textbf{-91}}$&\textbf{6.2}$^{\textbf{+1.5}}_{\textbf{-1.4}}$&\textbf{4.5}&  \textbf{-100} ($>$\textbf{400} \textbf{Hz}) &  \textbf{0-49} &  \textbf{1145} \\
\hline
 & 40030-01-04/06-00   &1310$^{+31}_{-33}$&208$^{+89}_{-61}$&3.4$^{+0.9}_{-0.8}$&4.2&  -100 &  All &   \\   
 &   &1282$\pm{27}$&223$^{+79}_{-58}$&3.7$\pm{0.8}$&4.9&  -75 ($>$400 Hz) &  0-49 &  \\   
 &   &\textbf{1285}$^{\textbf{+31}}_{\textbf{-32}}$&\textbf{249}$^{\textbf{+86}}_{\textbf{-63}}$&\textbf{3.8}\textbf{$\pm$0.8}&\textbf{4.9}&  \textbf{-100} ($>$\textbf{400} \textbf{Hz}) &  \textbf{0-49} &  \textbf{1185} \\ 
 \hline
  & 96404-01-10-01  &1266$^{+3}_{-10}$&23$^{+30}_{-20}$&2.5$^{+6.5}_{-0.7}$&3.6&  -100  &  All &  \\
   &   &1258$^{+8}_{-7}$&28$^{+45}_{-100}$&2.8$^{+39.8}_{-0.8}$&3.4&  -80  ($>$400 Hz) &  0-43 &  \\
  &    &\textbf{1269}$^{\textbf{+9}}_{\textbf{-14}}$&\textbf{51}$^{\textbf{+31}}_{\textbf{-39}}$&\textbf{2.6}$^{\textbf{+1.0}}_{\textbf{-0.8}}$&\textbf{3.1}&  \textbf{-100}  ($>$\textbf{400} \textbf{Hz}) &  \textbf{0-43} &  \textbf{1228} \\
\hline
4U\ 1728-34 &  50030-03-09-00/01  &1271$^{+15}_{-11}$&75$^{+37}_{-30}$&1.4$\pm{0.4}$&3.7&  -100 &  All &  \\ 
    &   &1283$^{+8}_{-13}$&51$^{+48}_{-148}$&1.1$^{+0.4}_{-0.3}$&4.1&  -75($>$400 Hz) &  0-43 &  \\
    &   &\textbf{1278}$^{\textbf{+15}}_{\textbf{-14}}$&\textbf{76}$^{\textbf{+42}}_{\textbf{-31}}$&\textbf{1.2}$^{\textbf{+0.4}}_{\textbf{-0.3}}$&\textbf{3.3}&  \textbf{-100} ($>$\textbf{400} Hz) &  \textbf{0-43}  & \textbf{1221}\\
\hline
 &  20083-01-02-000/01 &1277$^{+35}_{-42}$&238$^{+102}_{-66}$&1.3$\pm{0.3}$&4.3&  -100 &  All &  \\
  &   &1276$^{+35}_{-37}$&245$^{+102}_{-78}$&1.4$\pm{0.3}$&4.3&  -75($>$400 Hz) &  0-49 &  \\
  &   &\textbf{1280}$^{\textbf{+33}}_{\textbf{-41}}$&\textbf{254}$^{\textbf{+100}}_{\textbf{-81}}$&\textbf{1.4}\textbf{$\pm$0.3}&\textbf{4.8}&  \textbf{-100}($>$\textbf{400} \textbf{Hz}) & \textbf{0-49} &  \textbf{1156} \\
\hline
  & 20083-01-02-000 & 1293$^{+22}_{-35}$  & 127$^{+99}_{-60}$  & 1.7$^{+0.7}_{-0.6}$  & 2.9 & -100 & All\\
  &  & 1300$\pm{15}$  & 93$^{+59}_{-35}$  & 1.6$^{+0.6}_{-0.5}$  & 3.4 & -75 ($>$400 Hz) & 0-49 \\
    &  & \textbf{1302}$^{\textbf{+13}}_{\textbf{-10}}$  & \textbf{69}$^{\textbf{+54}}_{\textbf{-37}}$  & \textbf{1.4}$^{\textbf{+0.6}}_{\textbf{-0.4}}$  & \textbf{3.3} & \textbf{-100} ($>$\textbf{400} \textbf{Hz}) & \textbf{0-49}& \textbf{1213} \\
\hline 
4U\ 1636--53 & 60032-01-03-01(+30) & 1268$^{+5}_{-11}$  &53$^{+35}_{-33}$  & 0.24$\pm$0.06  & 4.0 & -100 & All & 
\\
  & & 1259$^{+8}_{-16}$  & 93$^{+120}_{-31}$ & 0.36$^{+0.18}_{-0.09}$ &  4.1 & -75($>$400 Hz) & 0-43 & \\  
 & & \textbf{1255}$^{\textbf{+13}}_{\textbf{-22}}$  & \textbf{134}$^{\textbf{+109}}_{\textbf{-61}}$ & \textbf{0.43}$^{\textbf{+0.15}}_{\textbf{-0.10}}$ & \textbf{4.4} & \textbf{-100} ($>$\textbf{400} \textbf{Hz}) & \textbf{0-43} & \textbf{1119} \\
\hline  
 SAX J1750.8--2900 & 60035-01-01-00 & 1245$^{+4}_{-8}$  &35$^{+35}_{-14}$  & 1.55$^{+0.7}_{-0.4}$ & 3.5 & -100 & All & \\
    & & 1250$^{+15}_{-8}$  & 63$^{+43}_{-31}$ & 1.63$^{+0.58}_{-0.46}$ &  3.6 & -75($>$400 Hz) & 0-43 & \\
     & & \textbf{1250}$^{\textbf{+9}}_{\textbf{-7}}$  & \textbf{54}$^{\textbf{+31}}_{\textbf{-25}}$ & \textbf{1.68}$^{\textbf{+0.53}}_{\textbf{-0.45}}$ & \textbf{3.8} & \textbf{-100} ($>$\textbf{400} \textbf{Hz}) & \textbf{0-43} & \textbf{1219} \\
     
\hline  
 4U\ 1246--59 & 90042-02-01/08/09-00 & 1258$^{+3}_{-2}$  &19$^{+10}_{-12}$  & 26$\pm3$ & 7.7 & -150 & All & \\
    & & 1255$^{+7}_{-4}$  & 6$^{+20}_{-31}$ & 40$^{+34}_{-36}$ &  1.1 & -75($>$400 Hz) & 0-43 & \\
     & & \textbf{1261}$^{\textbf{+0.9}}_{\textbf{-1.3}}$  & \textbf{16}$^{\textbf{+7}}_{\textbf{-9}}$ & \textbf{26}\textbf{$\pm$3} & \textbf{9.3} & \textbf{-200} ($>$\textbf{400} \textbf{Hz}) & \textbf{0-43} & \textbf{1256} \\
\hline
4U\ 1702--43 & 80033-01-10-00(+11) & 1198$^{+16}_{-14}$ &92$^{+55}_{-29}$  & 2.2$^{+0.7}_{-0.6}$ & 3.8 & -100 & All & \\
 &  & 1210$^{+27}_{-18}$ &154$^{+109}_{-55}$  & 3.2$^{+0.1}_{-0.1}$ & 3.9  & -75($>$400 Hz) & 0-43 & \\
 &  &  \textbf{1213}$^{\textbf{+28}}_{\textbf{-17}}$ & \textbf{154}$^{ \textbf{+109}}_{\textbf{-55}}$  &  \textbf{3.2}$^{\textbf{+0.1}}_{ \textbf{-0.1}}$ &  \textbf{4.0} &  \textbf{-100}($>$ \textbf{400}  \textbf{Hz}) &  \textbf{0-43} &  \textbf{1150}\\
 \end{tabular}
  }}
  \caption{Fit parameters of the upper kHz QPOs with the highest frequencies. We provide an estimate of the significance from the rms-normalized integral power divided by its negative error (IP/$\sigma^-$). Errors quoted here use $\Delta\chi^2$=1. In the last column we list the 3$\sigma$ (99.73$\%$) lower limit evaluated using $\Delta\chi^2$=7.74. Throughout this work we use the bold faced parameters. Regular type faced parameters are given for comparison only.}
  \label{fit}
\end{table*}

\subsection{EoS constraints}
\label{sec:eos}
The limits on mass and radius given in equations \ref{eq:M} and \ref{eq:R} are to first order in $j$. They are accurate for neutron stars with slow spin and have been recommended for use up to $ \nu_{\mbox{spin}}\approx$400 Hz \citep{Miller:1998}, although this spin limit depends on the equation of state. For faster spinning stars, the dependence of oblateness and internal structure of the neutron star on its rotation rate are not negligible, and their effect on the interior and exterior spacetime must be taken into account. Numerical methods are necessary to obtain accurate limits on mass and radius in these cases, as the test-particle geodesic equations need to be solved in the appropriate spacetime for a neutron star with a particular equation of state (EoS) \citep{Morsink:1999}. 

We compute the highest allowed orbital frequency for a star with a given EoS and spin frequency using the RNS\footnote{A code based on RNS designed to compute the ISCO is available at \url{https://github.com/rns-alberta/isco}} code by \cite{Stergioulas:1995} that follows the method by \cite{Cook:1992}. We choose 5 representative EoS that span a wide range of stiffness but also accommodate the 2.01$\pm$0.04$M_{\odot}$ mass of millisecond pulsar J0348+0432 \citep{Antoniadis:2013}. HLPS1, HLPS2, and HLPS3 are soft, intermediate and stiff versions of the realistic HLPS EoS \citep{Hebeler:2013} that describe the widest range of stiffness consistent with nuclear experiments. APR \cite{Akmal:1998} includes relativistic effects in the three-potential formalism, and L (reviewed in \citealt{Arnett:1977}) is based on mean-field theory \cite{Pandharipande:1975}.  The observation of the binary neutron star merger \citep{Abbott:2017}, most likely rules out EoS L, but it is still useful to include an unrealistically stiff EoS to demonstrate the effect of stiffness on the constraints found in this paper.

\begin{figure}
	\includegraphics[width=3in]{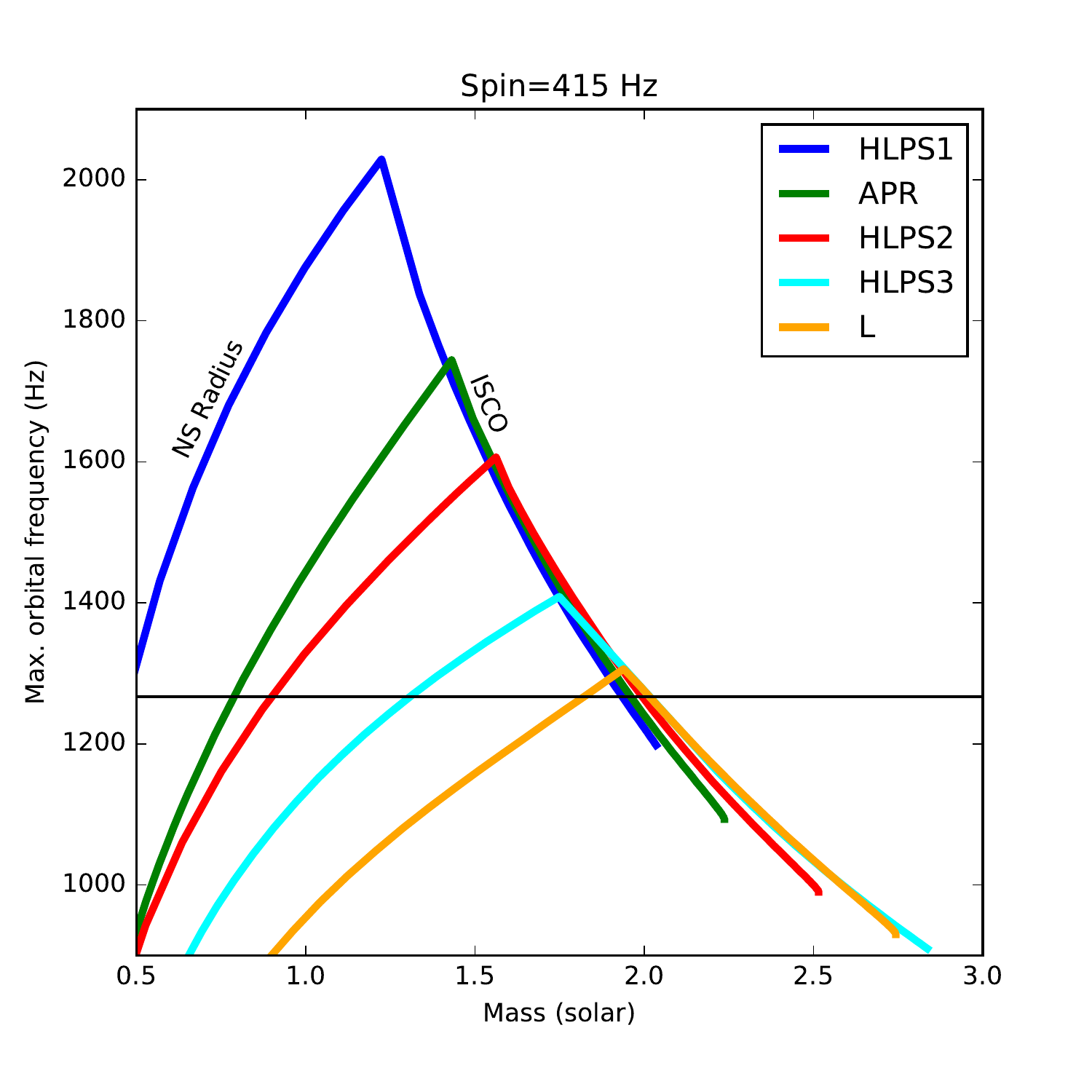}
   \caption{Highest possible circular orbital frequency vs. mass, around a neutron star with $ \nu_{\mbox{spin}}$=415 Hz for five EoS (HLPS1, HLPS2, HLPS3, APR and L). The highest 3$\sigma$ lower limit (1267 Hz) on the kHz QPO frequency for 4U 0614+09 ($ \nu_{\mbox{spin}}$=415 Hz) is plotted in black.}
   \label{fig:EoSMass}
\end{figure}

Given an EoS and a spin frequency, a range of stellar models can be computed, each with a different mass, radius, and largest orbital frequency. As an example, Figure \ref{fig:EoSMass} shows the largest orbital frequency for stars with different EoS spinning with a frequency of 415 Hz. In this figure, the largest orbital frequency for a particular EoS is represented by a curve with two branches. The low mass branch for each EoS, labelled ``NS Radius'', corresponds to stars that are larger than the ISCO radius, so that a test particle orbiting at the star's surface has the largest possible orbital frequency. The high mass branch for each EoS labelled ``ISCO'' corresponds to stars that are smaller than the radius of the ISCO. The maximum orbital frequency for a given EoS and spin corresponds to the frequency at the cusp where the two branches meet. On Figure \ref{fig:EoSMass} we show a  black horizontal line at 1267 Hz, the 3-$\sigma$ lower limit on the frequency of the most constraining QPO, 1267 Hz, for the neutron star 4U 0614+09 that spins at 415 Hz. If this QPO frequency corresponds to orbital motion, then it is clear that it is possible for all 5 representative EoS to explain the observed frequency. For a fixed value of spin, Figure \ref{fig:EoSMass} also shows that the ISCO-frequency-branch curves for the different EoS follow approximately the same curve. All of the ISCO frequency curves intersect the 1267 Hz line at a value close to 2.1 $M_\odot$, indicating an EOS-independent maximum possible mass for 4U 0614+09 if the orbital frequency interpretation of the upper frequency QPO is adopted. Since the QPO could correspond to orbital motion at distances further from the star than the ISCO, smaller masses are allowed. In the case of the two stiffest EoS, the QPO frequency also provides a lower mass limit of 1.3 $M_\odot$ for EoS HLPS3 and 1.8 $M_\odot$ for EoS L. We assume that it is unphysical for neutron stars with masses smaller than 0.8 solar mass to exist, so there is no lower mass limit on the softer EoS given by this QPO interpretation.  The observation of a higher frequency QPO from this neutron star has the potential to rule out the stiffer EoS in our sample. 

Similar upper mass limits on the other neutron stars can be found by making similar plots of largest orbital frequency as a function of mass for the appropriate spin frequency. The resulting maximum masses are obtained by assuming the 3-$\sigma$ limit on the observed upper frequency QPO is orbital in nature. We find a mass constraint for 4U 1702--43 ($<$2.2 M$_{\odot}$), SAX J1750.8-2900 ($<$2.3 M$_{\odot}$), 4U 1636--53 ($<$2.4 M$_{\odot}$), 4U 1728--34 ($<$2.1 M$_{\odot}$) using this interpretation. Since this limit is set by the ISCO radius, it is robust in the sense that its dependence on the EoS is weak.

A stronger assumption is that the highest frequency QPO corresponds to motion of a test particle at the ISCO radius. In the case of 4U 0614+09 the neutron star's mass would be required to be $2.0 \pm 0.1 M_\odot$ with the uncertainty coming from the unknown EoS.  

\begin{figure}
	\includegraphics[width=3in]{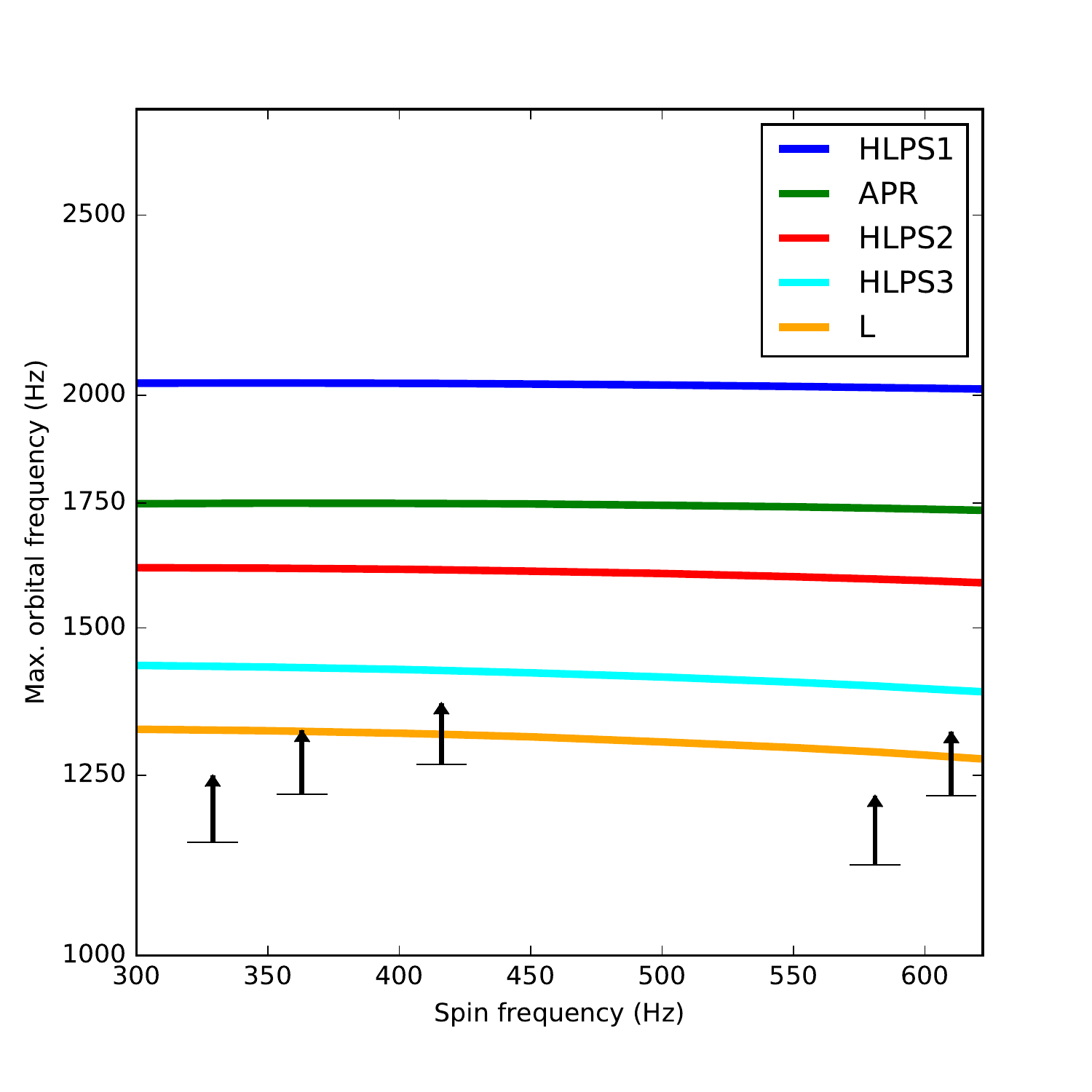}
   \caption{Highest possible circular orbital frequency vs. spin frequency for stable neutron stars of any mass, for five EoS (HLPS1, HLPS2, HLPS3, APR and L, see text).  The highest 3$\sigma$ lower limits on the kHz QPO frequency for 4U 1728--34 ($ \nu_{\mbox{spin}}$=363 Hz), 4U 0614+09 ($ \nu_{\mbox{spin}}$=415 Hz), 4U 1636-53 ($ \nu_{\mbox{spin}}$=581 Hz), SAX J1750.8--2900  ($ \nu_{\mbox{spin}}$=601 Hz) and 4U 1702-43 ($ \nu_{\mbox{spin}}$=329 Hz) are overplotted. Orbital motion is a viable explanation for the kHz QPO for each of these EoS.}
   \label{fig:EOS}
\end{figure}

The maximum allowed orbital frequencies for the five representative EoS are shown as a function of spin frequency in Figure \ref{fig:EOS}. The largest $3\sigma$ lower limits on the upper kHz QPO frequency are shown on Figure \ref{fig:EOS} for the five neutron stars with known values of spin. As can be seen from the figure, all these EoS allow orbital frequencies consistent with the 3$\sigma$ lower limits on kHz QPO centroid frequencies we obtain, which supports the hypothesis that kHz QPO frequencies are orbital and constrained by neutron star and ISCO radii.\\

\begin{figure}
	\includegraphics[width=3in]{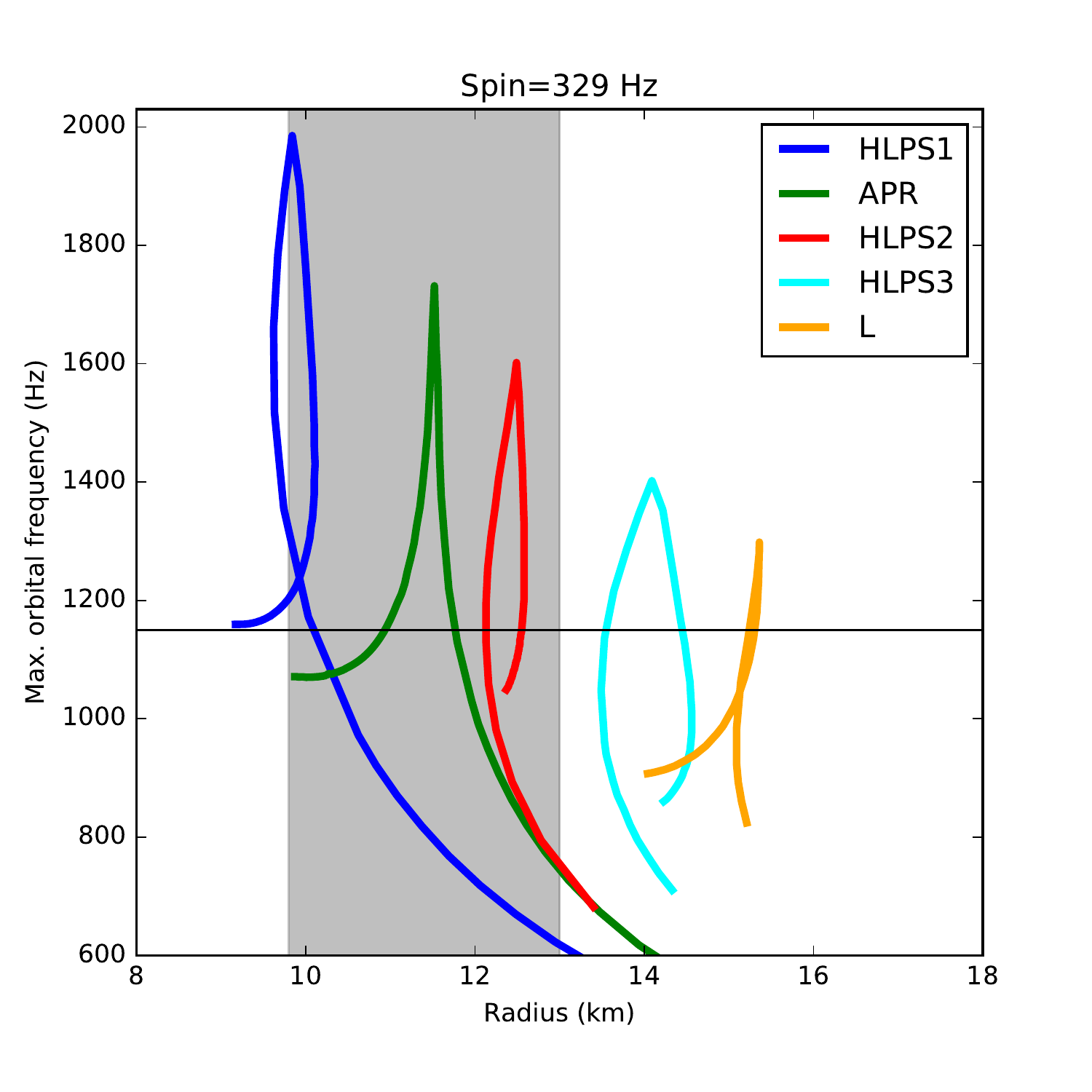}
   \caption{Highest possible circular orbital frequency vs. radius, around a neutron star with spin=329 Hz for five EoSs (HLPS1, HLPS2, HLPS3, APR and L, see text). The highest 3$\sigma$ lower limit (1150 Hz) on the kHz QPO frequency for 4U 1702-43 is plotted in black. The 12.4$^{+0.6}_{-2.6}$ km (3$\sigma$) radius constraint obtained by \protect\cite{Nattila:2017} is indicated by the shaded area.}
   \label{fig:EoSR}
\end{figure}

In Figure \ref{fig:EoSR} we plot the highest possible orbital frequency vs. radius for these EoS for an assumed $ \nu_{\mbox{spin}}$ of 329 Hz, the spin frequency of 4U 1702--34. 

The lower limit to the highest observed frequency QPO, 1150 Hz, is plotted as a black horizontal line. If this QPO represents orbital motion, then for a given EoS the allowed properties of 4U 1702--34 must be given by a value that lies above the horizontal line on the appropriate curve. This restricts the possible range of radii for any particular EoS for this star to fairly narrow ranges ($\pm$0.1--$\pm$0.7 km depending on the EoS).
A radius constraint of 12.4$^{+0.6}_{-2.6}$ (3$\sigma$) was recently obtained by \cite{Nattila:2017}. This range of radii would only allow the softer EoS: APR, HLPS1 and HLPS2.  For these EoS, if the kHz QPO is generated at the ISCO and under the reasonable assumption that NS masses are $>$0.8 M$_{\odot}$, the NS mass in 4U\ 0614+09 would be 2.0$\pm$0.1. 
However, we caution that the constraints found by \cite{Nattila:2017} assumes spherical, non-rotating stars so it is not clear how to compare their radius constraint with our graph.

We also checked if our kHz QPO frequencies, if orbital, provide further constraints on the specific moment of inertia, a quantity that is important, for instance, in the relativistic precession model \citep{Stella:1998}. We estimated $\emph{I}_{45}\emph{m}^{-1}$ (where $M=m$$\cdot$$M_{\odot}$ and $I=10^{45}I_{45}$ g cm$^{2}$ are the neutron star mass and moment of inertia, respectively) from our most constraining case of the 3$\sigma$ lower limit of 1267 Hz around a neutron star with 415 Hz spin in 4U 0614+09. We find ranges for L of (1.8--1.9), for HLPS3 of (1.3--1.7), for HLPS2 of (0.9--1.36), for HLPS1 of (0.4--0.94) and for APR of (0.8--1.1). Overall, for these EoS, a range of 0.4--1.9 is allowed, in accordance with the range 0.5-2.0 commonly used \citep{Friedman:1986}. \footnote{For the 1219 Hz lower limit in SAX J1750.8--2900 ($ \nu_{\mbox{spin}}$=601 Hz) we find ranges for L of (1.9--2.0), for HLPS3 of (1.4--1.9), for HLPS2 of (1.0--1.4), for HLPS1 of (0.46--2), and for APR of (0.8--1.1).}


\subsection{Lower kHz QPO}

In \cite{Boutelier:2009} a drop in Q of the lower kHz QPO (L$_{\ell}$) as $\nu_u$ increases above 1000 Hz is reported. This behaviour is explained by a QPO-generating region of the accretion disk approaching the ISCO. The upper kHz QPO represents the local orbital frequency. It is expected that when the quality factor drops to zero, the ISCO and therefore the maximum kHz QPO frequency is reached. Extrapolating the declining trend in Q$_{\ell}$ vs. $\nu_{\ell}$ toward zero yields a highest $\nu_{\ell}$ of 920-930 Hz. Using the frequency difference (320 Hz) between the lower and upper kHz QPOs, this interpretation allows a maximum upper kHz QPO of $\sim$1250 Hz. Our 1267 Hz lower limit somewhat exceeds this maximum QPO frequency and is therefore relevant in this context. 

\subsection{Conclusion}

We investigated reports of $>$1300 Hz kHz QPOs in RXTE archival data on NS-LMXBs that could possibly tighten constraints on the mass and radius of neutron stars. We find that a 2-18 keV energy selection improves the detection significance of $>$1000 Hz QPOs. Using this selection, we obtain precise measurements of the highest kHz QPO frequencies in 6 neutron stars. 
We detect a kHz QPO at $\nu_u$=1288$\pm$8 Hz with a 3$\sigma$ lower limit of 1267 Hz in 4U\ 0614+09. This constitutes our best-constrained case. For the five sources in our sample with known spin, we compare our lower limits to the predicted maximum possible circular orbital frequency around a NS calculated in full GR taking into account the effect of their known spin on stellar structure and spacetime for five relevant EoS. In the most constraining case of the 1267 Hz 3$\sigma$ lower limit, we find an upper limit on the NS mass of M$<$2.1 M$_{\odot}$ 
For 4U\ 1702--43, we find that soft to intermediately stiff EoS (HLPS1-2, APR) are able to explain the recent radius constraint obtained by modelling the spectra of X-ray burst cooling tails \citep{Nattila:2017}. In combination with the highest 3$\sigma$ lower limit on the kHz QPO in that source, the radius is constrained to a narrow range different for each EoS. Assuming the kHz QPO is generated at or just outside the ISCO, the mass of the NS can be obtained, which for 4U 0614+09 is 2.0$\pm$0.1 M$_{\odot}$.
We conclude that orbital motion around the NS constrained by stellar and ISCO radii is a viable explanation for the kHz QPO for each of the EoS included in this work. 

\section*{Acknowledgements}
This research has made use of data obtained through the High Energy Astrophysics Science Archive Research Center Online Service, provided by the NASA/Goddard Space Flight Center. This work is (partly) financed by the Netherlands Organisation for Scientific  (NWO). NSERC provided funding for SMM's research. We thank the referee for the constructive feedback and suggestions that improved the quality of the paper.
\\




\bibliographystyle{mnras}
\bibliography{Bibliography} 








\bsp	
\label{lastpage}
\end{document}